\documentclass[aps,prx,cha,graphicx,amsmath,amssymb,reprint,onecolumn,superscriptaddress,floatfix]{revtex4}
\usepackage{graphicx}

\begin{document}

\title{How crosslink numbers shape the large-scale physics of cytoskeletal materials}
\author{Sebastian F\"urthauer}
\affiliation{Center for Computational Biology, Flatiron Institute, New York, NY 10010, USA} 
\author{Michael J. Shelley}
\affiliation{Center for Computational Biology, Flatiron Institute, New York, NY 10010, USA}
\affiliation{Courant Institute, New York University, New York, NY 10012, USA}

\begin{abstract}
    Cytoskeletal networks are the main actuators of cellular mechanics, and a
    foundational example for active matter physics. In cytoskeletal networks,
    motion is generated on small scales by filaments that push and pull on
    each other via molecular-scale motors. These local actuations give rise to
    large scale stresses and motion. To understand how microscopic processes
    can give rise to self-organized behavior on larger scales it is important to
    consider what mechanisms mediate long-ranged mechanical interactions in the
    systems. Two scenarios have been considered in the recent literature. The
    first are systems which are relatively sparse, in which most of the large
    scale momentum transfer is mediated by the solvent in which cytoskeletal filaments are
    suspended. The second, are systems in which filaments are coupled via crosslink
    molecules throughout. Here, we review the differences and commonalities
    between the physics of these two regimes. We also survey the literature for
    the numbers that allow us to place a material within either of these two
    classes.
\end{abstract}

\maketitle
\section{Introduction}
Living systems move. On cellular scales this motion is actuated by networks of
polymer filaments crosslinked by molecular-scale motors that exert
forces between them. This system is collectively referred to as the
cytoskeleton \cite{howard01, alberts07}. Understanding how the material properties of the
cytoskeleton
emerge from the properties of cytoskeletal components is one of the
great challenges for soft condensed matter physics and cell biology
\cite{foster2019cytoskeletal}. Solving it
will allow biologists to predict the effects of molecular scale perturbations on
cellular organelles and enable physicists and engineers to pursue
strategies for developing similarly complex and robust materials in the lab.
Here, we take measure of the current state of theory for understanding
the emergent physics of cytoskeletal systems based on filament scale
interactions.

Symmetry based theories have helped clarify the fields, such as densities,
velocities and order parameters, through which the dynamics of active
materials can be described on long length and time scales \cite{kruse2005generic,
furthauer2012active, julicher2018hydrodynamic}. These theories have also 
provided means for generating a more or less complete list of material properties
(or phenomenological coefficients), such as viscosities, elastic
moduli, and activity coefficients which characterize a material. While this type
of information can be used to relate the effects of molecular perturbations
to material properties \cite{naganathan2018morphogenetic}, they do not address
the question how micro-scale processes set the large-scale physics.

To approach this challenge physicists and mathematicians have generalized the
methods developed for polymeric and liquid crystalline systems
\cite{doi1988theory, degennes1995}, and Boltzmann's statistical physics for
gases, to incorporate the effects of the microscopic activity of molecular-scale
motors. These efforts have led to a set of theories which describe relatively
sparsely connected polymeric assemblies whose large scale behavior is dominated
by momentum transfer via solvent flows \cite{liverpool2003instabilities, aranson2005pattern,
gao2015multiscalepolar}. In parallel, other groups have extended the physics of
elastic networks to obtain theories for the physics of
crosslinked materials on time scales that are short compared to crosslinker
dynamics \cite{broedersz2014modeling}.

However, recent work has increased the awareness that many cytoskeletal systems,
belong to a third class of active materials. They are highly crosslinked gels,
in which filaments are transiently coupled by a large number of motors and
crosslinks, which bind and unbind on time scales that are fast compared to the
long time, large scale dynamics of the system. This highly crosslinked regime is set apart
by new physical phenomena. Most strikingly, filaments in highly crosslinked
systems can slide through the gel at speeds that are independent of the
structure of their local
surrounding. This phenomenon has been predicted theoretically
and observed {\it in vitro} \cite{furthauer2019self}, and {\it in vivo}
\cite{mitchison1989polewards, yang2008regional, dalton2021gelation}. Theories for highly crosslinked systems take
inspiration from early work on actin bundles
\cite{kruse2000actively,kruse2003self} and
describe larger networks that are crosslinked by molecular motors with different
symmetries and force velocity relations 
\cite{furthauer2021design}.

The goal of this review is to give a perspective on our current understanding of
the physics of cytoskeletal materials. We highlight the differences between
highly and sparsely crosslinked cytoskeletal networks, and their models.  Where
possible from published data, we will classify known examples along these
lines. Along the way we will identify and highlight some open challenges that
the field needs to address in order to enable a quantitative and predictive
physics of these living materials.

In this review, we focus on  continuum models and the microscopic models
underlying their derivations. We do however emphasize that symmetry based
theories \cite{marchetti2013hydrodynamics}, continuum models obtained by coarse
graining microscopic models \cite{saintillan2013active, maryshev2018kinetic},
and agent based models for the same systems
\cite{janson2007crosslinkers,head14,popov16,freedman17} have been extensively
studied in numerical simulations as well.

In section \ref{sse:cytoskeleton} we review the key constituents of cytoskeletal
Then, in section
\ref{sse:classify} we propose a classification of cytoskeletal structures as
either highly or sparsely crosslinked. To guide us in this, we review the
conditions under which a materials long-range momentum transport is dominated by
crosslink interactions or by solvent flow, respectively. We then classify
studied cytoskeletal materials given published data.  After that, in Section
\ref{sse:theory}, we review the theoretical descriptions of both highly and
sparsely crosslinked materials, highlighting and contrasting their different
physics. We point out how these differences can be used both as a predictive and
diagnostic tools.

\section{Key constituents of the cytoskeletal networks}\label{sse:cytoskeleton}
The cytoskeleton is the cellular machinery which enables cells to do mechanical
work on their organelles and on their surroundings.
\begin{figure}[h]
    \centering
    \includegraphics[width=\columnwidth]{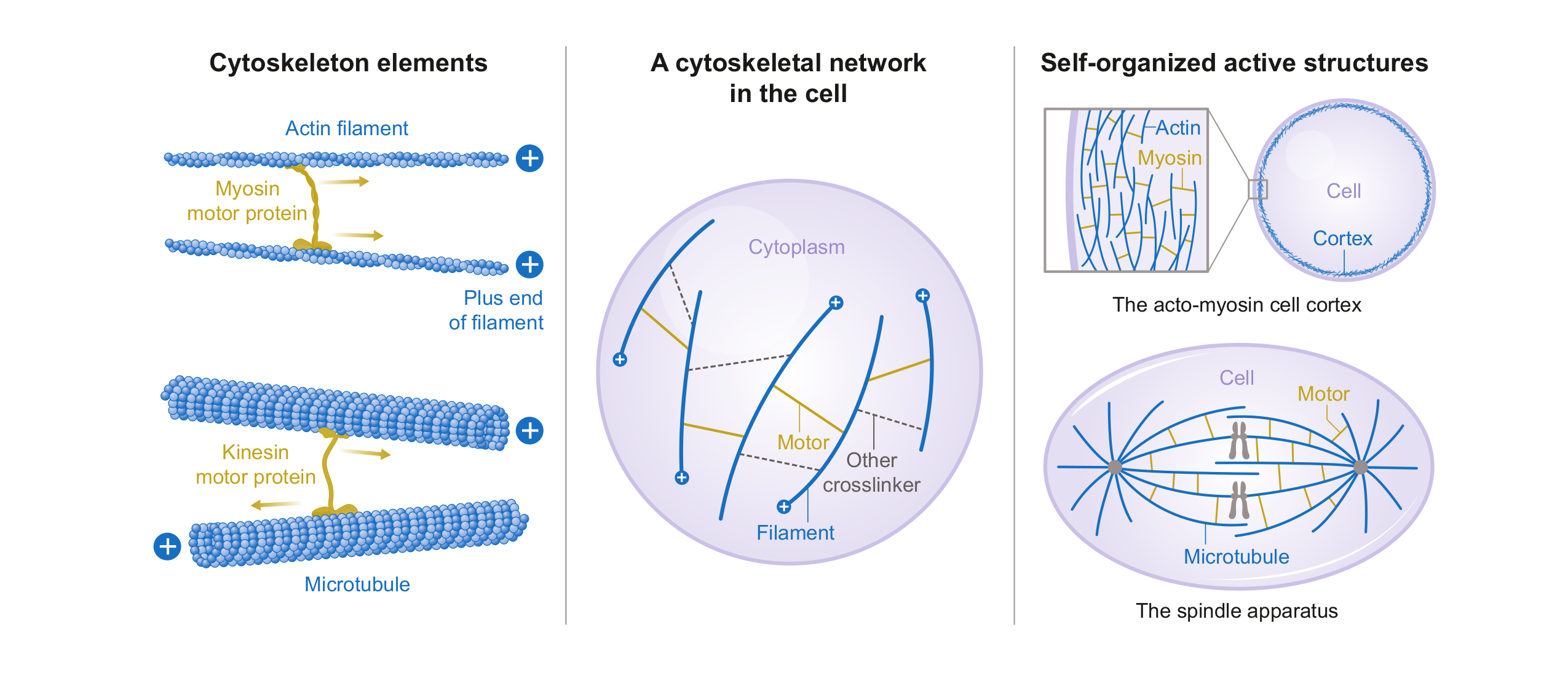}
    \caption{Polar cytoskeletal filaments are coupled and actuated by
      molecular scale motors (left). Collectively they form the
      cytoskeletal networks (middle) which are the materials that make
      up many important biological structures (right).}
    \label{fig:fig1}
\end{figure}
The cytoskeleton consists of polymer filaments, proteins that crosslink and
actuate these filaments, and regulators which help organize and coordinate their
functions in time and space. All of these constituents are suspended in
cytoplasm, which is the aqueous slurry that makes up a cell's interior.

\subsection{Cytoskeletal Filaments}
Several types of cytoskeletal filaments exist. The two most studied and most
abundant ones are filamentous actin  and microtubules \cite{alberts07}; see
Fig.~{(\ref{fig:fig1})}. Both 
are structurally polar, meaning that their molecular structure differentiates
one direction from the other. They are 
both transient since they are constantly assembling and
disassembling. They are active since filament assembly and disassembly are driven by
energy derived from GTP. While for our purposes here, it is sufficient to think
of cytoskeletal filaments as polar rods, we want to emphasize that in typical
{\it in vivo} systems the lifetime of actin filaments and microtubule is often
much shorter than the time scale of large scale rearrangements of the system.
One typical example is gastrulation in fruitfly. This process, that is driven by
actomyosin contractions, takes about 2 hours and is driven by actomyosin, which
turns over within minutes \cite{collinet2021programmed}.
Another such process is the separation of chromosomes during cell division in
the nematode worm {\it C. elegans},
which takes tens of minutes and is achieved by microtubules that have lifetimes
of tens of seconds \cite{redemann2017celegans}. There are  additional
cytoskeletal filaments beyond actin and microtubules, largely called intermediate filaments.
Examples include keratins, which are a large family of proteins that encompass
vimentins, which play a role in cell motility, and lamins, which
act as a mechanical scaffolding for the cell nucleus.  These
intermediate filaments - so-called since their thickness in the electron
tomographs where they where first characterized was between that of thin
actin filaments and that of thick Myosin mini-filaments - are typically
longer-lived, apolar, and are hypothesized to provide cells with long time
elasticity.  Kinesins, myosins and dyneins that link intermediate filaments to
the microtubule and actin cytoskeleton have been identified, but so far their
role remains under-explored.

\subsection{Active and passive crosslinkers}
The second class of molecules that are important for the cytoskeleton are
crosslinking proteins which connect cytoskeletal filaments to each other and
to other structures such as chromosomes and membranes. Too many types exist to do them justice in a
review focussed on the physics of the cytoskeleton, so we refer the reader to
\cite{pollard2016actin} and \cite{alfaro2015building} for recent perspectives
on actin and microtubule
associated crosslinkers. Many of these crosslinkers are molecular
scale motors. This means they have access to a chemical fuel reservoir - in
general Adenosine-Tri-Phosphate (ATP) - from which they can draw power to do
mechanical work upon the
structures they crosslink. Motors can walk along cytoskeletal filaments, acting
as moving crosslinks, which 
slides filaments past each other. Or, they can carry a cargo, like a vesicle or
a mitochondrion, along filaments.
The most abundant actin-associated molecular motors are myosins. The most
abundant microtubule-associated ones are dyneins and kinesins. Typically motors
can exert forces of up to several pico-Newtons, and their unloaded walking
speeds on filaments can vary from nanometers to microns per second. Finally, motors can
be more or less processive. A motor's processivity measures the expected number
of steps a motor typically takes, before it unbinds from the filament to which
it is attached \cite{howard01,alberts07}. 

\subsection{Cytoplasm}
The last important actor in understanding the physics of the cytoskeleton is the
aqueous slurry in which it is immersed. While the rheology of the cytoplasm is most
certainly complex \cite{luby1999cytoarchitecture}, most current physical
theories for the cytoskeleton ignore this fact - mainly because the detailed
characteristics of cytoplasmic rheology are poorly understood and difficult to
model. The theories that
we review, thus approximate it as a very viscous Newtonian
fluid with a viscosity between 100 and 1000 times that of water; see for example
\cite{garzon2016force}.

\subsection{Forces and torques in cytoskeletal networks}
\begin{figure}[h]
    \centering
    \includegraphics[width=\columnwidth]{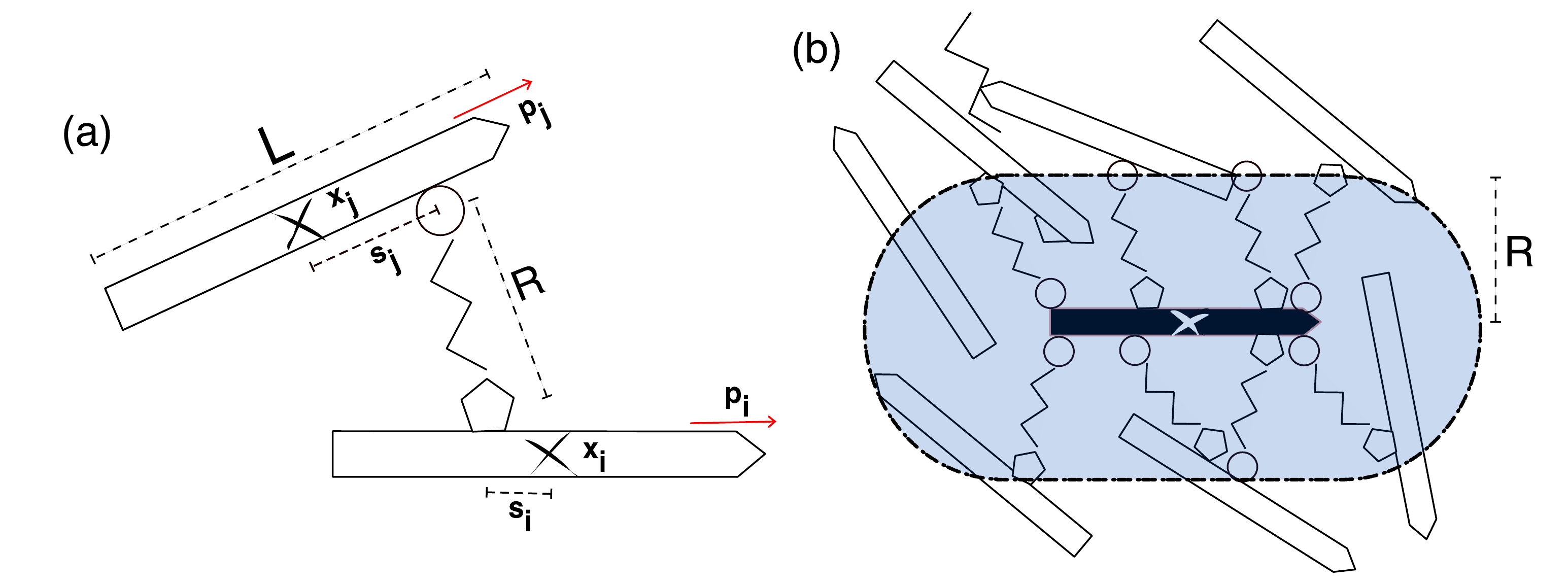}
    \caption{ (a) Interaction between two cytoskeletal filament $i$ and
      $j$ via a molecular motor.  Filaments are characterized by their
      positions $\mathbf x_i, \mathbf x_j$, their orientations
      $\mathbf p_i, \mathbf p_j$, and connect by a motor between
      arc-length position $s_i, s_j$. A motor consist of two heads
      that can be different (circle, pentagon) and are connected by a
      linker (black zig-zag) of lengt $R$ (b) The total force on
      filament $i$ is given by the sum of the forces exerted by all
      $a$ (circle) and $b$ (pentagon) heads, which connect the
      filament into the network. The shaded area shows all
      geometrically accessible positions that can be crosslinked to
      the central (black) filament.}
\label{fig:cartoons}
\end{figure}
The goal of a physical theory for the cytoskeleton is to predict how filaments
in a cytoskeletal network rearrange in response to the forces that they
experience and produce. The larger goal is a quantitative understanding of the
mechanical processes that allow cytoskeletal structures to self-organize and
perform their complex tasks within cells.

The elementary starting point is characterizing the
total force $\mathbf F_i$ that acts on the $i$-th filament in the network. We assume
that it has the form
\begin{equation}
    \mathbf F_i = \sum_j\mathbf F_{ij}^{\times} + \mathbf F_i^{s} = 0,
\end{equation}
which vanishes since cytoskeletal systems are overdamped; see
Fig.~{\ref{fig:fig1} (middle)}.
The total force on each filament consists of crosslink mediated forces
$\sum_j\mathbf F_{ij}^{\times}$ and a contribution $\mathbf F_{i}^{s}$, which comes
from the drag between filaments and a moving solvent. The crosslink mediated force
between filaments $i$ and $j$ is given by 
\begin{equation}
    \mathbf F_{ij}^{\times} =
    \int_{-L/2}^{L/2}ds_i\int_{-L/2}^{L/2}ds_j\int_{\Omega(\mathbf x_i
    +s_i\mathbf p_i)}d^3x~\delta(\mathbf x -\mathbf x_j -s_j \mathbf
        p_j)\mathbf f_{ij},
\end{equation}
where $\mathbf f_{ij}$ is the force density exerted by crosslinkers between the
arclength positions $s_i$ and $s_j$ on the 
two filaments $i,j$, respectively; see Fig.~\ref{fig:cartoons}.
Here, $\Omega(\mathbf x)$ is a sphere
centered around the point $\mathbf x$ and whose radius $R$ is the size
(maximal extension) of the cross-linker. Here all
filaments are taken to have the same length $L$. 
Analogous expressions for the torques and total torque on filament $i$ hold.

In the following we will argue that modeling of cytoskeletal networks has
focussed on two different regimes: (i) a highly crosslinked regime in which the
long range momentum transport through the gel is dominated by crosslinking
forces and (ii) a regime in which momentum transport through the solvent
dominates. Theories for the two limits make very different predictions for the
scaling of important material properties and dynamics.

\section{Which cytoskeletal materials are highly crosslinked?}
\label{sse:classify}
\begin{figure}[h]
    \centering
   \includegraphics[width=\columnwidth]{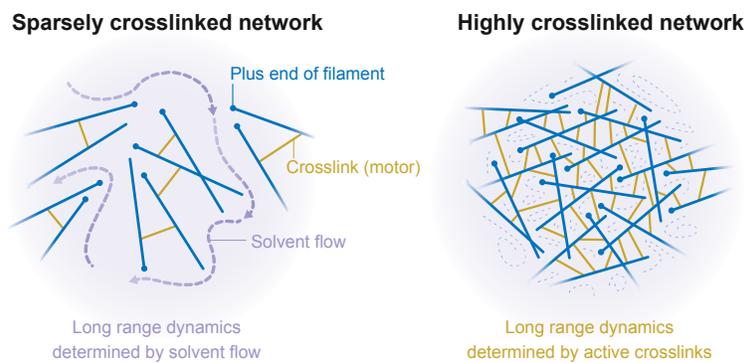}
   \caption{Illustration of the key differences between highly and sparsely
   crosslinked networks, emphasizing the dominant mechanisms of large scale
   momentum transport.}
   \label{fig:fig_sparse_dense}
\end{figure}
We now discuss the crossover between the highly and sparsely crosslinked
regimes and try to classify a few well studied cytoskeletal systems along these lines. For this
we define  the density of cytoskeletal filaments $\rho=\sum_i \delta(\mathbf x_i -\mathbf x)$ and 
the force density $ f =\sum_i \delta(\mathbf x_i -\mathbf x) \mathbf
F_i$. Since crosslink mediated interaction are short ranged we can write
\begin{equation}
    \mathbf f = \nabla\cdot\mathbf \Sigma - \Gamma \rho (\mathbf v-\mathbf
    v^{s}) +\mathcal O\left( L^4 \right) = 0,
    \label{eq:force_gel}
\end{equation}
where $L$ is assumed small compared to the
system scale \cite{furthauer2021design}.  In the following we use
the word gel to signify the filaments and crosslinks and solvent for the
cytoplasm
in which they are immersed. 
The gel stress $\mathbf \Sigma$ encodes the momentum flux through
the crosslinks which connect filaments, and $\Gamma\rho$ is the friction coefficient between the gel and the
solvent. Finally, $\mathbf v$ and $\mathbf v^s$ are the
gel and solvent center of mass velocities, respectively.
If we take the solvent to be a Newtonian fluid with viscosity $\mu$ it obeys
the Stokes equation
\begin{equation}
    \mu\Delta\mathbf v^{s} -\nabla \mathbf \Pi^s + \Gamma \rho (\mathbf
    v-\mathbf v^{s}) = 0,
    \label{eq:force_solvent}
\end{equation}
where $\mathbf \Pi^s$ is the hydrostatic pressure. The characteristic length
scale over which
momentum is transported through the solvent, the permeation length, is thus given by
\begin{equation}
    \ell^s = \sqrt{\frac{\mu}{\rho\Gamma}},
\end{equation}
which decreases as the density $\rho$ of filaments in the
gel increases. Conversely, crosslinks generate friction between the
filaments they connect, which leads to a
viscous contribution $ \mu^g \rho^2
\nabla \mathbf v$ to $\mathbf \Sigma$, i.e. proportional the strain rate  of the
gel and to the number of filament-filament interactions mediated by crosslinks
\cite{furthauer2019self,furthauer2021design}. Here $\mu^g\rho^2$
is the gel viscosity.  Thus we expect the typical length scale over which
momentum is transported through the gel to be   
\begin{equation}
    \ell^g = \sqrt{\frac{\mu^g\rho^2}{\Gamma\rho}},
\end{equation}
which increases with filament density. 
The implied assumption of this calculation is that the number of
realized crosslink connections is limited by the number of geometrically possible
filament-filament interactions and not the number of crosslinking molecules available. It can
however easily be generalized. 

In the following we consider a system highly crosslinked if $\ell^g \gg \ell^s$
and sparsely crosslinked if $\ell^s \gg \ell^g$.  We next review the literature
and overview which cytoskeletal systems are highly or sparsely crosslinked,
respectively; see Fig.~\ref{fig:fig_sparse_dense}.

\subsection{The cell cortex}
\begin{figure}[h]
    \centering
    \includegraphics[width=\columnwidth]{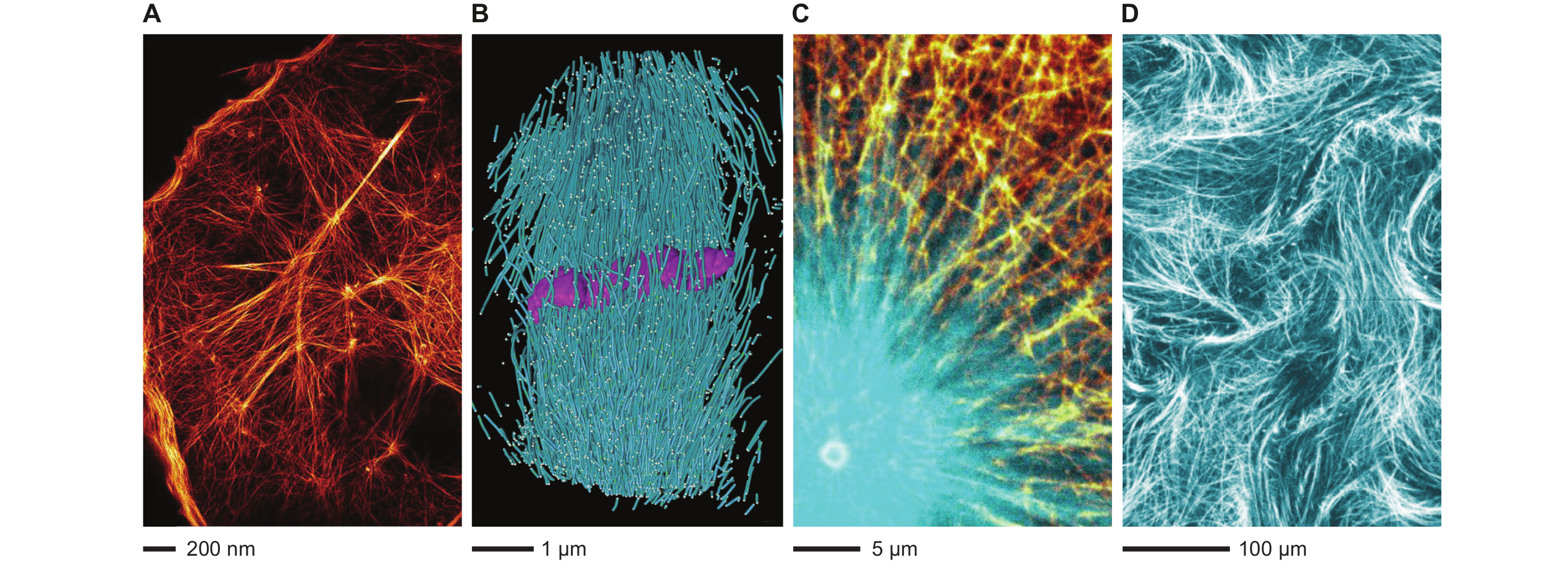}
    \caption{Biological and artificial cytoskeletal networks: (A)
      SIM-TIRF super-resolution microscopy image of the actin cortex
      from a kidney culture cell. With permission from
      \cite{li2015extended}.  (B) Electron tomography reconstruction
      of the inner spindle microtubules of a embryonic {\it
        C. elegans} mitotic spindle. Taken from
      \cite{redemann2017celegans}; (C): spinning disk fluorescence
      microscopy image of microtubules and actin in a {\it Xenopus
        laevis} interphase aster in egg extract. Courtesy of James
      Pelletier; (D) snapshot from an experiment an active microtubule
      nematic synthesized from engineered multimeric kinesin-1 motors
      and stabilized microtubules. Adapted with permission from
      \cite{wu2017transition}. Throughout the figure actin filaments a
      represented in orange and microtubules in blue.  }
    \label{fig:fig_exp}
\end{figure}
The cell cortex is a thin layer of actin filaments that localize at the cell
membrane; see Fig.~{\ref{fig:fig_exp} (A)}. It is actuated by non-muscle myosin 2 and held together by a variety
of crosslinkers and proteins which regulate assembly and disassembly of both
thick (myosin) filaments and thin (actin) filaments \cite{pollard2016actin}. 
Typically the thickness of the cortical layer is 300-1000nm 
\cite{salbreux2012actin}. Flow
of actomyosin is vital for cell polarity establishment (the process by which cells
break the symmetry between two daughter cells)
\cite{mayer2010anisotropies,gross2019guiding},
cell migration and shape changes \cite{callan2016cortical},  cell division
\cite{turlier2014furrow}, as well as regulating 
cellular surface tension \cite{chugh2017actin}. On time-scales longer than
filament lifetimes, the cell cortex is well described as an 
active fluidic  material \cite{mayer2010anisotropies, naganathan2014active,
naganathan2018morphogenetic}.

Comprehensive estimates of the numbers of crosslinks in the actin cell cortex
are hard to come by. That said, many actin filaments are nucleated by
the nucleator Arp2/3 which is a crosslinker itself. In
\cite{fritzsche2016actin} the authors estimate that roughly nine out of ten actin
filaments are nucleated by Arp2/3, while about one out of ten is nucleated by
formins. 
This finding suggests that the actin cortex is
highly crosslinked starting from its nucleation. Moreover, actin filaments are
actively crosslinked by myosin motors and other crosslinkers. In
\cite{wu2005counting} the authors count the numbers of the most
important actin binding partners in fission yeast cells. They find that the number of
crosslinkers and motors acting on actin filaments is comparable (or larger) than the number
of nucleators (Arp2/3 and formins) in the system. Thus, if one
estimates that the number of filaments is roughly proportional to the number of available
nucleators, one concludes that there are several crosslinking interactions per filament
exist at any give time. That is, the actin cortex is highly crosslinked.

\subsection{The spindle}
The spindle is the cellular apparatus which segregates chromosomes during cell
division. Its main constituents are microtubules which are crosslinked and actuated
by dynein and kinesin molecular motors; see Fig.~{\ref{fig:fig_exp} (B)}.
Despite having common constituents and function, spindle structure can vary
widely, even in a single organism. 

One spindle whose physics has been studied extensively is the meiotic spindle
of {\it Xenopus laevis}.
This spindle can be reconstituted in frog egg extract, which makes it particularly
attractive to study {\it ex vivo}. These spindles are large - about 50 microns long -
and consist of microtubules whose lengths and lifetimes are exponentially
distributed around an average length and lifetime of about 6
microns and 20 seconds, respectively \cite{brugues2012nucleation}. They have been shown to obey the fluctuation spectra
expected for active nematic gels \cite{brugues2014physical}. In these spindles,
microtubules nucleate by an auto-catalytic mechanism in which pre-existing
microtubules recruit growth factors that initiate the formation of new
microtubules \cite{petry2011augmin, petry2013branching,oh2016spatial,
kaye2018measuring, decker2018autocatalytic}. This mechanism implies that filaments are created
crosslinked to the spindle gel. The details of the pathway controlling nucleation
have been investigated \cite{thawani2018xmap215, thawani2019spatiotemporal}. 
The motor proteins dynein and Eg-5 kinesin crosslink the spindle network
further. While direct measurements of the number of Eg-5 kinesin crosslinks are
not available,
observations of microtubule motion in spindles allow an estimate. In
\cite{dalton2021gelation} the authors quantify the motion and polarity of
microtubules in {\it Xenopus} spindles. They estimate that two or motors
per filament are required to explain experimental observation that velocity
fluctuation are highly correlated throughout the whole spindle. In our language
this means that $\ell^g$ is
comparable to system size and thus that the {\it Xenopus} spindle is highly
crosslinked.

Another well studied spindle is the embryonic mitotic spindle of the
nematode worm {\it C. elegans}. These spindles are much smaller - 
$10\mu$m pole to pole - and have been fully reconstructed using electron
tomography \cite{redemann2017celegans}; see Fig.~\ref{fig:fig_exp} (C). Using this structural information,
together with light microscopy and mathematical modelling, 
 the authors in \cite{redemann2019current} argue that the connection between
spindle poles and chromosomes is indirectly maintained by microtubules
crosslinking to other microtubules. While the evidence is indirect it is
suggestive of a highly crosslinked network.  

The third example that we want to highlight here is the female meiotic spindle,
also in {\it C . elegans}. This spindle is about half the size of the mitotic
one and has also been reconstructed in tomography. The findings from tomography
studies \cite{redemann2018switch, lantzsch2020microtubule} suggest that it
consists of short and short-lived microtubules and is structurally quite similar
to the {\it Xenopus laevis} meiotic spindle. It is thus tempting to speculate
that this spindle too is a highly crosslinked network.

\subsection{Microtubule asters}
In cells, spindles are positioned for cell division by microtubule asters which
emanate from the spindle poles and grow out towards the cortex; see
Fig.~\ref{fig:fig_exp} (C).  In smaller
cells, microtubules interacting with cortex are important for spindle
positioning
\cite{pavin2012positioning,garzon2016force,nazockdast2017cytoplasmic}.  Recent
work in {\it C. elegans} embryos suggests that dynein mediated pulling forces
from cortically bound dynein are crucial \cite{farhadifar2020stoichiometric}.

In larger cells, such as the {\it Xenopus} zygote, the situation is somewhat
more complicated, since cytoskeletal filament are in general much shorter than
the typical cell radii \cite{sulerud2020microtubule}. Thus it has been proposed
that in these cells spindle are positioned by cytoplasmic pulling, that is by
dynein motors which carry a cargo from the aster periphery towards its center.
This would cause an active drag against the cytoplasmic fluid, which would ultimately
cause spindle centering \cite{meaders2020microtubule,
xie2020cytoskeleton}.  To test this hypothesis the authors of
\cite{pelletier2020co} co-imaged the cytoskeletal actin, microtubules, and
membranous networks. They find that these networks move against each other near
aster boundaries, but move together near the asters center. It is thus tempting
to ask what explains these different behaviors. One appealing hypothesis is that
these asters are highly crosslinked near their centers, and less crosslinked
near aster cores.

\subsection{Artificial systems}
Given the complexity and the sheer number of different constituents of
cytoskeletal networks in living cells, biophysicists have developed
simplified {\it in vitro} systems made from cytoskeletal
components; see Fig.~\ref{fig:fig_exp} (C). Here, the physics of
cytoskeletal networks can be studied at a reduced complexity and in a more controllable
environment. Beyond that, they have provided foundational examples of
out-of-equilibrium materials.

One prominent system by  \cite{sanchez2012spontaneous} is made from stable
microtubules and kinesin-1 motors joined by engineered linkers. 
This same mixture has been used to study active materials in bulk and on 2d
interfaces \cite{sanchez2012spontaneous},
in vesicles \cite{keber14}, rigid confinements \cite{opathalage2019self,
chandrakar2020confinement} and droplets \cite{vcopar2019topology}.  The same
system has also been adapted to study disclination loops in 3d active nematics
\cite{duclos2020topological}.  It is surprising, given the importance of
these systems, that the structure and number of crosslinks that bind microtubules
remain relatively unexplored. More studies will be needed to solidly classify
these particular systems as either highly or sparsely crosslinked.

Somewhat more recently, actin-based active networks have been synthesized.  Many
of the key findings of this line of research a summarized in
\cite{banerjee2020actin}.  These systems are in general driven by the molecular
motor myosin and require the admixture of passive crosslinkers to generate
active contractions. This peculiarity has been attributed to a liquid-gel
transition, which may imply that the contractile states of these actomyosin
networks are in the highly crosslinked regime. 

There is a growing list of {\it in vitro} systems, using different motors and
filaments assembled in cell extract or fully reconstituted such as in
\cite{foster2015active, foster2017connecting} or the ones reviewed in
\cite{needleman2017active,banerjee2020actin}.  For most of these systems it
remains unclear where they fall on the range from sparsely to highly connected
systems.

\section{Theoretical descriptions of cytoskeletal systems}
\label{sse:theory}

\begin{figure}[h]
    \centering
    \includegraphics[width=\columnwidth]{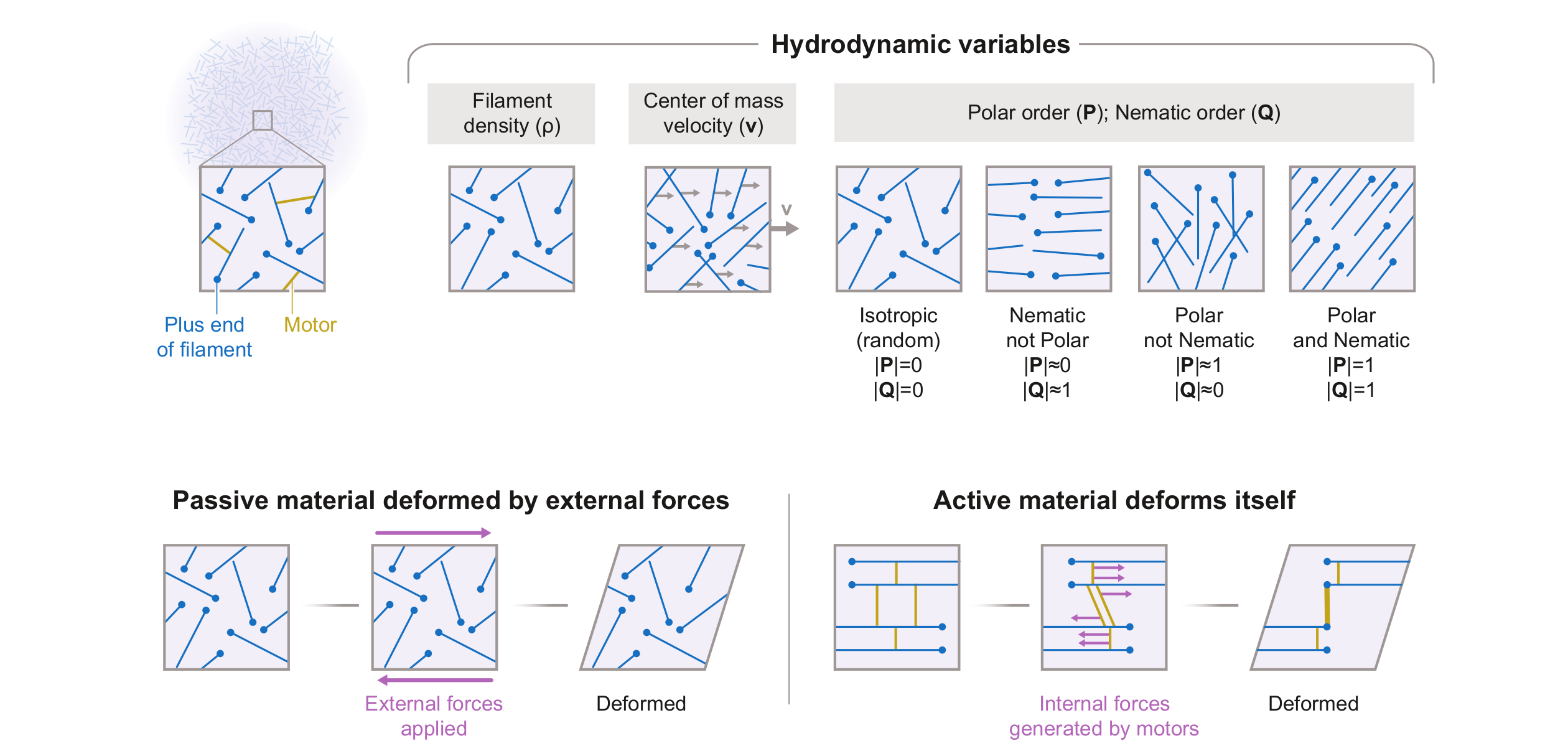}
    \caption{Top: Illustration of the important hydrodynamic variables. 
    Bottom: Illustration of externally strained and self-strained materials. }
    \label{fig:fig_hydro}
\end{figure}

Much of our current theoretical understanding of cytoskeletal active
matter derives from the remarkable success of extending the broken-symmetry
hydrodynamics of liquid crystal system \cite{degennes1995} to active
materials where micro-scale non-equilibrium processes can generate force and
torque dipoles \cite{kruse2005generic, furthauer2012active,
julicher2018hydrodynamic}.
These theories have clarified the fields (densities, velocities, order
parameters) through which to describe the dynamics of a cytoskeletal material
on long-wave hydrodynamic scales; see Fig.~{\ref{fig:fig_hydro}}, and which 
material properties, such as the viscosities and activity
coefficients, are needed to characterize these materials. 
More microscopic
theories derive the values of material properties and their microscopic origins.

We start by reviewing the fields and material properties that we use to
characterize cytoskeletal materials. We then review theoretical predictions for
how these material properties are determined in highly and sparsely crosslinked
networks.

\subsection{Hydrodynamic variables}
We seek to describe the evolution of a cytoskeletal system with a
small number of continuous fields. At large length and time scales the fields
needed to sufficiently characterize a material are those associated  with
conserved quantities and those associated to continuous broken symmetries
\cite{chaikin2000principles}; see Fig.~\ref{fig:fig_hydro} (top).
We first introduce the fields associated to conserved quantities.

The first is the total mass density
$\rho^\mathrm{tot}(\mathbf x,t)$ of the material. Since mass can be
neither created nor destroyed,
\begin{equation}
    \partial_t\rho^\mathrm{tot} = -\nabla\cdot\mathbf g^\mathrm{tot},
\end{equation}
where $\mathbf g^\mathrm{tot}(\mathbf x,t)$, denotes the total mass flux - or
momentum density. In the cytoskeletal systems of interest here the
total mass density consists of solvent and
gel contributions, such that $\rho^\mathrm{tot}=\rho+\rho^\mathrm{s}$,
where $\rho$ is the gel contribution, by which we mean the mass of
filaments and proteins attached to filaments.  In general, chemical
processes, such as filament nucleation and polymerization,
and the binding and unbinding of proteins, can convert solvent mass into
gel mass and vice versa. Thus we can write
\begin{eqnarray}
    \partial_t\rho &=& -\nabla\cdot\rho \mathbf v + R
    \\
    \partial_t\rho^\mathrm{s} &=& -\nabla\cdot\rho^\mathrm{s}\mathbf v^\mathrm{s} -
    R,
\end{eqnarray}
where $R$ is the mass conversion rate from chemical reactions
and we have defined
the gel and solvent velocities by $\mathbf g^\mathrm{tot} = \rho \mathbf v + \mathbf
 \rho^\mathrm{s} \mathbf v^\mathrm{s}$.
In living
materials the dominant contribution to $R$ is filament polymerization and
depolymerization, which in general is highly regulated. For instance in {\it
Xenopus}'s meiotic spindles the dominant mode of filament creation is autocatalytic,
and new filaments nucleate in a branching process from preexisting ones
\cite{petry2011augmin, petry2013branching,oh2016spatial, kaye2018measuring,
decker2018autocatalytic,thawani2018xmap215, thawani2019spatiotemporal}.

The second field that arises from conservation considerations is the momentum
density itself. In the absence of external forces, the momentum density 
$\mathbf g^\mathrm{tot}(\mathbf x,t)$ can only change by a momentum flux
\begin{equation}
    \partial_t \mathbf g^\mathrm{tot} = \nabla\cdot\mathbf \Sigma^{tot},
    \label{eq:force_balance}
\end{equation}
where $\mathbf\Sigma^{tot}(\mathbf x,t)$ is total stress in the material.
Like the mass flux we can write $\mathbf\Sigma^{tot} = \mathbf\Sigma +\mathbf\Sigma^{s}$, and
 force balance equations for the gel
and solvent components of the material separately. Under the assumptions that
the solvent is Newtonian,  that inertial terms can be neglected, and
that the force exerted between the gel and fluid phases is linear in the gel
density and their velocity difference, one recovers Eqns.~{(\ref{eq:force_gel},\ref{eq:force_solvent})}. 
Thus, the fields that we need to keep track of, since they are
associated to conserved quantities, are $\rho, \rho^\mathrm{s},\mathbf
v$, and $\mathbf v^\mathrm{s}$. We next introduce the fields that
are associated with continuous broken symmetries.

Order parameter fields characterize the conformational state of the
material. For the purposes here we assume that cytoskeletal filaments are
rod-like and rigid.
Then the tensor orientation order parameter
$\mathcal{Q}(\mathbf{x},t)$ naturally arises. From $\mathcal Q$
devolves the scalar orientation order parameter and director
field. Unlike purely nematic systems, cytoskeletal filaments like
microtubules or actin are polar, as are the motor-mediated forces
acting upon them. Hence, we will also need a polar order
parameter $\mathbf P(\mathbf x,t)$. If the system were characterized
at the microscopic level by a particle distribution function
$\psi(\mathbf x,\mathbf p,t)$ \cite{doi1988theory} with $\mathbf p$ being the
particle orientation vector on the 2-sphere, then these order
parameters arise as its $\mathbf p$-moments $\mathbf P=\left<\mathbf p
\right>/\rho$ and $\mathcal Q=\left<\mathbf p \mathbf p
\right>/\rho$, where the angular brackets denote a distributional averaging over
the $\mathbf p$-sphere.
In general higher order symmetry fields
exist, but we ignore them here for simplicity. 

In summary, we seek to describe the cytoskeletal systems in terms of
just a few coarse grained continuous fields. They are the densities
$\rho$ and $\rho^s$, the velocity fields $\mathbf v$ and $\mathbf v^s$,
and the order parameter fields $\mathcal Q$ and $\mathbf P$. Thus, the next task
is to arrive at expression for the stresses $\mathbf \Sigma^\mathrm{tot}$ and
$\mathbf\Sigma$, and the transport operators for the order parameter fields
$\mathcal Q$ and $\mathbf P$, and the chemical reaction rate $R$, in terms of
hydrodynamic variables. In this review we are concerned with the
expressions for the stresses first and foremost.

\subsection{Material properties of active cytoskeletal networks}
Given the set of hydrodynamic variables, symmetry-based theories postulate
material laws for all unknown quantities by writing down all symmetry allowed
terms that respect the invariances of the systems, such as translation and
rotation invariance. How to obtain symmetry based
material equations from symmetry considerations or from non-equilibrium
thermodynamics considerations is reviewed elsewhere in great detail
 \cite{marchetti2013hydrodynamics,julicher2018hydrodynamic}.

For our purposes here, it is sufficient to express the total stress by the equation, 
\begin{equation}
    \mathbf \Sigma^\mathrm{tot} = \chi^v:\nabla\mathbf v + \chi^s:\nabla\mathbf v^s + \chi^a:\nabla\mathbf P 
    + \Pi^a \mathcal I 
    + \mathcal A^{(\mathbf P)}\mathbf P\mathbf P
    - \mathcal A^{(\mathcal Q)}\mathcal Q 
    + \bar{\mathbf \Sigma},
    \label{eq:stress_generic}
\end{equation}
which obeys the symmetry constraints discussed in \cite{marchetti2013hydrodynamics,julicher2018hydrodynamic}.
Here, $\chi^v$ is the fourth-rank gel viscosity tensor, $\chi^s$ is a fourth-rank
solvent viscosity tensor, and $\chi^a$ is a
fourth-rank tensor that quantifies active polar stresses proportional to
gradients in gel polarity $\mathbf P$.  The coefficients $\mathcal A^{(\mathbf
P)}$ and $\mathcal A^{(\mathcal Q)}$ quantify the magnitude of active polar and
active
nematic stresses in the material, respectively.
Furthermore $\Pi^a$ acts as an active pressure and
$\bar{\mathbf \Sigma}$ denotes other stresses in the system,
which stem from hydrostatic and steric exclusion and alignment effects.
For an in-depth discussion see
\cite{degennes1995} and \cite{julicher2018hydrodynamic}.  
Together, $\chi^v, \chi^a, \chi^s, \Pi^a, \mathcal A^{(\mathbf P)}$, and $\mathcal
A^{(\mathcal Q)}$ are the material properties of the cytoskeletal filament
network. They in turn depend on the state of the network, that is the
hydrodynamic variables, and need either be measured in an experiment, or
derived from more microscopic considerations to allow a complete description of
the material.
Symmetry-based theories also elucidate the form of the transport equations  for $\mathbf P$ and
$Q$. We refer to
\cite{marchetti2013hydrodynamics,julicher2018hydrodynamic} for a more complete
discussion. 

In this review we are concerned with theories that derive expressions
for the phenomenological coefficients in
(Eq.~{(\ref{eq:stress_generic})}) from micro-scale considerations of
motors and filaments. In particular we want to contrast theories for
highly crosslinked and sparsely crosslinked systems. We will highlight
that the functional dependencies of $\chi^v, \chi^a, \chi^s, \Pi^a,
\mathcal A^{(\mathbf P)}$ and $\mathcal A^{(\mathcal Q)}$ change
between these two regimes.

\subsection{The highly crosslinked regime of cytoskeletal networks}
We first review theories for the hydrodynamic limit of the highly
crosslinked regime of cytoskeletal networks. We consider a
cytoskeletal material to be highly crosslinked if $\ell^g \gg
\ell^s$. The hydrodynamic limit of such a material is dominated by
crosslink interactions, and thus solvent mediated
interactions can be ignored, since the length scale of momentum
transport associated with $\mathbf \Sigma$ is much larger than that
associated with $\mathbf \Sigma^\mathrm{s}$; see
Fig.~\ref{fig:fig_sparse_dense}. Thus these theories assume that
$\mathbf \Sigma^\mathrm{tot} = \mathbf \Sigma$ which implies $\mathbf v^s =
\mathbf v$ and that $\mathbf \Sigma^\mathrm{s}$ is negligible.

To our knowledge this limit of active gels was first discussed in the context of
studying actin bundles \cite{kruse2003self, kruse2003continuum} where the
authors proposed a 1d model for describing actin
structures in the cortex. Only recently have these ideas been extended to larger
three-dimensional systems.  In \cite{furthauer2019self, furthauer2021design} a
phenomenological model for the force density $\mathbf f_{ij}$, generated by
crosslinks between filaments, was proposed. In this model the force density is
given by
\begin{eqnarray}
    \mathbf f_{ij} &=& K(s_i, s_j)\left( \mathbf x_i +s_i\mathbf p_i -
    \mathbf x_j -s_j\mathbf p_j\right) 
    \nonumber\\
    &+& \gamma (s_i, s_j)\left( \mathbf v_i +s_i\dot{\mathbf p}_i - \mathbf
    v_j -s_j\dot{\mathbf p}_j\right)
    \nonumber\\
    &+& \left[ \sigma(s_i, s_j) \mathbf p_i - \sigma(s_j, s_i) \mathbf p_j \right ].
    \label{eq:generic_force_density}
\end{eqnarray}
The functions $K(s_i, s_j),~ \gamma(s_i,s_j),~ \sigma(s_i,s_j)$ characterize the
types of forces exerted by crosslink populations between the arclength positions $s_i, s_j$ on
filaments $i$, and $j$. The coefficient $K$ captures effects of crosslink
elasticity, the coefficient $\gamma$ acts as an effective frictional coupling
between filaments, and $\sigma$ characterizes the active forcing from motor
stepping; see Fig.~\ref{fig:cartoons}. The main conceptual advance between this
phenomenological model and the ones proposed in earlier work, is that here
$\mathbf f_{ij}$ explicitly depends on the velocities and rates of rotation of
filaments $i,j$. With this dependence, in a dense system the dynamics of
all filaments in the system becomes tightly coupled. In contrast, earlier work
\cite{kruse2000actively, aranson2005pattern, LM2006, maryshev2018kinetic} had
posited motor interactions  that depend only on the relative
orientations and center of mass positions of filaments, effectively constraining themselves to
pairwise interactions.  

From Eq.~(\ref{eq:generic_force_density}) one calculates the stress in the network using
\begin{eqnarray}
    &&\mathbf \Sigma =  
    -\frac{1}2 \sum_{i,j}
    \left\lfloor
    \begin{array}{c}
    \delta(\mathbf x-\mathbf x_i)\delta(\mathbf x^\prime-\mathbf x_j)
    \\ 
\left(\mathbf x_i -\mathbf x_j +s_i \mathbf p_i - s_j \mathbf p_j\right)\mathbf f_{ij}
\end{array}
\right\rceil_{\Omega(\mathbf x_i)}^{ij}
    \label{eq:rod_stress}
\end{eqnarray}
where the brackets $\lfloor\cdots\rceil^{ij}_{\Omega(\mathbf x_i)}$
denotes the coarse graining operation
\begin{equation}
    \lfloor\phi\rceil_{\Omega(\mathbf x_i)}^{ij} = \int\limits_{-\frac{L}{2}}^{\frac{L}{2}} ds_i
    \int\limits_{-\frac{L}{2}}^{\frac{L}{2}} ds_j
    \int\limits_{\Omega(\mathbf x_i)}d\mathbf x^3 \phi. 
    \label{eq:bracket_def}
\end{equation}
Here  $\Omega(\mathbf x)$ is a sphere
centered around the point $\mathbf x$ and whose radius $R$ is the size
(maximal extension) of the cross-linker. In this version of the theory, all
filaments are taken to have the same length $L$. Thus, the
operation $\lfloor\cdots\rceil^{ij}_{\Omega(\mathbf x_i+s_i\mathbf
  p_i)}$ integrates its argument over all geometrically possible
crosslink induced interactions.
    
The emergent  material properties turn out to depend
on a small set of $s$-moments of the phenomenological coefficients $K,~
\gamma,~ \sigma$. These moments are given by
\begin{equation}
    X_{nm}(\mathbf x) =  \lfloor X(s_i, s_j) s_i^n s_j^m\rceil_{\Omega(\mathbf x)}^{ij},
    \label{eq:moment_definition}
\end{equation}
where $X$ can be $K,~ \gamma,$ or $\sigma$.  More explicitly,
\begin{eqnarray}
    K_0  &=& K_{00} = \left\lfloor K(s_i, s_j)\right\rceil_{\Omega(\mathbf x_i)}^{ij}, 
    \\
    K_1  &=& K_{10} = K_{01} =\left\lfloor s_i K(s_i, s_j)\right\rceil_{\Omega(\mathbf x_i)}^{ij}, 
    \\
    \gamma_0  &=& \gamma_{00} =\left\lfloor \gamma(s_i, s_j)\right\rceil_{\Omega(\mathbf x_i)}^{ij}, 
    \\
    \gamma_1  &=& \gamma_{10} = \gamma_{01} = \left\lfloor s_i \gamma(s_i, s_j)\right\rceil_{\Omega(\mathbf
    x_i)}^{ij}, 
    \\
    \sigma_0  &=& \sigma_{00} = \left\lfloor\sigma(s_i, s_j)\right\rceil_{\Omega(\mathbf x_i)}^{ij}, 
    \\
    \sigma_{10}  &=& \left\lfloor s_i \sigma(s_i,
    s_j)\right\rceil_{\Omega(\mathbf x_i)}^{ij}, 
    \\
    \sigma_{01}  &=& \left\lfloor s_j \sigma(s_i,
    s_j)\right\rceil_{\Omega(\mathbf x_i)}^{ij}. 
\end{eqnarray}
With these definitions, one finds for the viscosity tensor
\begin{equation}
    \chi_{\alpha\beta\gamma\mu}^v = \rho^2\gamma_0\left(\frac{3R^2}{10}\delta_{\alpha\gamma}\delta_{\beta\mu}
    +
    \frac{L^2}{12}\gamma_0\left(\mathcal Q_{\alpha\gamma}\delta_{\beta\mu} \mathcal
- Q_{\alpha\beta}\mathcal Q_{\gamma\mu}\right)\right),
    \label{eq:passive_viscosity}
\end{equation}
and, very importantly, that
\begin{equation}
    \chi^a_{\alpha\beta\gamma\mu} =
    \frac{\sigma_0}{\gamma_0}\chi_{\alpha\beta\gamma\mu}^v.
    \label{eq:self_strain_visc}
\end{equation}
for the coefficient quantifying the active polar stress.

The relation between Eqs.~{(\ref{eq:passive_viscosity}) and
  (\ref{eq:self_strain_visc})} implies that we can combine the first two terms
  of Eq.~\ref{eq:stress_generic} into 
  \begin{equation}
      \mathbf\Sigma^{ss} = \chi^v:\left(\nabla \mathbf v
      +\frac{\sigma_0}{\gamma_0}\nabla\mathbf P  \right)
  \end{equation}
  which we call the self-straining stress.
  In this type of material $\sigma_0/\gamma_0$ is essentially
the free velocity of the motors and crosslinks coupling filaments.
Thus the state of zero self-straining stress is one where the viscous stress exactly
balances the active polar stress, which can be achieved when all filaments in the
material move at the motor speed in the direction in which they point \cite{furthauer2019self}. 
This self-straining state 
 is a consequence of the
fact that the same active crosslinks which drive filament motion also couple
the dynamics over long length scales. In essence, in a material where the
crosslinking is provided by active crosslinks, filaments can move
through the material at the preferred speed of the motor without
stressing the material; see Fig.~\ref{fig:fig_hydro}.

The same calculation also makes predictions for other active stresses in the
material \cite{furthauer2021design}. One finds
\begin{eqnarray}
    A^{(\mathcal Q)} &=& \rho^2(\sigma_{10}-\sigma_0\frac{\gamma_1}{\gamma_0}
    +\frac{L^2}{12}K_0), 
    \label{eq:AQ}
    \\
    A^{(\mathbf P)} &=& \rho^2(\sigma_{01}-\sigma_0\frac{\gamma_1}{\gamma_0}),
    \label{eq:AP}
    \\
    \Pi^a &=& \rho^2 \frac{3R^2}{10} K_0.
    \label{eq:Pi}
\end{eqnarray}
These expressions generalize to the continuum limit the predictions made by
\cite{belmonte2017theory} in the context of a discrete model.

An analogous calculation provides
expressions for the filament speed $\mathbf v_i$ and rotation rate $\dot{\mathbf
p}_i$ in the lab frame: 
\begin{equation}
    \mathbf v_i -\mathbf v = \frac{\sigma_0}{\gamma_0}(\mathbf p_i -\mathbf P)
+\mathcal O{(L^2)}
    \label{eq:vi_dense_limit}
\end{equation}
and
\begin{eqnarray}
    \dot{\mathbf p}_i &=& \left( \mathcal{I} -\mathbf p_i \mathbf p_i
    \right)\cdot
    \left\{
        \begin{array}{c}
            (\nabla\mathbf v + \frac{\sigma_0}{\gamma_0}\nabla\mathbf P )\cdot\mathbf p_i 
        \\+ \frac{12}{\gamma_0L^2\rho^2}\mathcal E\cdot\mathbf p_i 
        \\+
    \frac{12 }{\gamma_0L^2}A^{(\mathbf P)} \mathbf P 
\end{array}
\right\},
    \label{eq:kinetic_angular_dense}
\end{eqnarray}
where $\mathcal E$ is the steric distortion field \cite{degennes1995}. We refer
the reader to \cite{furthauer2021design} for additional detail.

\subsection{The sparsely crosslinked regime of cytoskeletal networks} 
We now review theories of the sparsely crosslinked regime of
cytoskeletal networks. We consider a cytoskeletal material to be
sparsely crosslinked if $\ell^s \gg \ell^g$, so that the hydrodynamic
(long-wave) limit is dominated by fluid mediated interactions and the
viscous transport of momentum though the gel is negligible; see
Fig.~\ref{fig:fig_sparse_dense}.  Thus,
these approaches proceed from Eq.~(\ref{eq:force_solvent}).  Given a
microscopic model of how the gel deforms in the fluid, these models
proceed to calculate the active stresses $\mathbf \Sigma^a$, typically
reflecting force dipoles acting on the solvent. Thus,
$\Gamma\rho(\mathbf{v}-\mathbf{v}^s)\simeq\nabla\cdot\mathbf\Sigma^a$. This
implies $\mathbf{\Sigma}^\mathrm{tot}=
\mathbf{\Sigma}^\mathrm{s}+\mathbf\Sigma^\mathrm{a}$. The motion of
the gel generally takes the form $\mathbf v^s -\mathbf v \propto
\mathbf P$, which allows polar pieces of gel to 'swim' through the
solvent.

One such line of models devolves from the mechanics of swimming
suspensions. To study the dynamics of suspensions of hydrodynamically
interacting swimming microorganisms, Saintillan \& Shelley
\cite{saintillan2008instabilitiesPRL,saintillan2008instabilitiesrPHYSFLUID} developed a kinetic theory coupling a
Smoluchowski equation for particle position and orientation to a
forced Stokes equation; see also \cite{SK2009,BM2009,saintillan2013active}. At the
particle level, free swimmers (which are force dipoles in
the solvent) are treated as rod-like particles swimming in a locally
linear flow, propelled by a prescribed active surface stress. The
swimmer velocity, rotation rate, and entire surface stress, can be
calculated in terms of the background flow, with the background flow
itself determined through solution of a forced Stokes equation. The
forcing is through an ``extra stress'' found using Batchelor's
adaptation \cite{Batchelor1970}, for dilute suspensions, of Kirkwood
theory. In its simplest formulation, Saintillan \& Shelley find the
coupled Smoluchowski equation for the particle distribution function
$\psi(\mathbf{x},\mathbf{p},t)$
\begin{equation}
  \partial_t\psi+\nabla_x\cdot(\dot{\mathbf{x}}\psi)
  +\nabla_p\cdot(\dot{\mathbf p}\psi)=0, \textrm{  where  }
  \dot{\mathbf{x}}=\mathbf{v}^s(\mathbf x)+U_0\mathbf p \textrm{  and  }
  \dot{\mathbf p}=(\mathbf I - \mathbf{p}\mathbf{p})
  \nabla\mathbf{v}^s(\mathbf{x})\mathbf{p},
  \label{Eq:SS1}
\end{equation}
where $\mathbf{p}$ is swimmer orientation and $U_0$ is its undisturbed
speed, and the last expression is Jeffrey's equation ({\it cf.}
Eq.~{(\ref{eq:vi_dense_limit},\ref{eq:kinetic_angular_dense})}). The solvent velocity
$\mathbf{v}^s$ is found through solution of a Stokes equation driven by
an active dipolar stress $\mathbf{\Sigma}^a$:
\begin{equation}
  -\nabla_x\Pi^{s}+\mu\Delta\mathbf{v}^s=-\nabla_x\cdot\mathbf{\Sigma}^a
  \textrm{  and  }\nabla_x\cdot\mathbf{v}^s=0,
  \textrm{  with  }\mathbf{\Sigma}^a=\alpha_0\rho\mathcal{Q}.
  \label{Eq:SS2}
\end{equation}
Here, $\alpha_0$ is the so-called dipole strength whose sign reflects
whether the swimmer force dipole is extensile ($\alpha_0<0$) or
contractile ($\alpha_0>0$). This categorization depends upon the
detailed placement of thrust and no-slip regions on the effective
swimmer body. This theory predicts both the instability of aligned
suspensions \cite{SR2002,ESS2013}, and the instability of isotropic
extensile suspensions when swimmers exceed a critical concentration
\cite{saintillan2008instabilitiesrPHYSFLUID,HS2010,ESS2013}. In later work \cite{ESS2013},
inter-particle aligning torques and consequent stresses, based on
Maier-Saupe theory \cite{doi1988theory}, were introduced to capture
sterically-induced concentration effects; see \cite{SS2013} for a
review.

Gao {\it et al.} \cite{gao2015multiscalepolar,GBGBS2015b} adapted this
approach to treating immersed assemblies of microtubules undergoing polarity
sorting by multimeric motor complexes such as the engineered kinesin-1
complexes used in \cite{sanchez2012spontaneous}. To do this, as the
basic unit they considered nematically ordered local clusters of
microtubules, say with $n$ of them pointing rightwards and $m$
pointing leftwards (hence, $\mathbf{P}=(n-m)/(n+m)\hat{\mathbf{x}}$);
see Figure 1 of \cite{gao2015multiscalepolar}. It is assumed that
every microtubule pair in the cluster is connected by active, plus-end
directed cross-linkers moving at speed $v_w$ on each microtubule.  The
coupling between the anti-aligned populations induces a minus-end
directed relative sliding for each.  By using Stokesian slender-body
theory \cite{KR1976} and assuming that the solvent drag thus calculated
is the only force resisting microtubule motion they calculated the
velocities of the left- and rightward pointing MTs to be
$v_L=(2n/n+m)v_w$ and $v_R=-(2m/n+m)v_w$. This yields
$v_R-v_L=-2v_w$. Slender-body theory again yields the forces each
rod exerts upon the fluid, and hence the induced “extra stress” tensor
arising from polarity sorting within the bundle can be calculated and
is found to be proportional to $v_w mn/(m+n)\hat{\mathbf x}\hat{\mathbf x}$.

Thus, in a cluster with $m=n$, the two populations are pulled past
each other with equal speed $v_w$, while producing a maximal
polarity-sorting stress. If most microtubules point rightwards, so
that $m\approx 0$, then $v_L\approx 2v_w$ and $v_r\approx 0$, and the
polarity sorting extra stress is small (with a like statement if $n\approx 0$). 
That the microtubule speeds depends upon the polarity is a
consequence of the micro-mechanical model wherein only solvent drag
resists microtubule translocation.

Adapting this cluster picture to a more general setting where the
cluster also moves with the background fluid velocity, Gao {\it et
  al.} give the analogous result to Eqs.~(\ref{Eq:SS1},\ref{Eq:SS2}) where the
microtubule flux and active stress tensor, induced by polarity sorting
of anti-aligned microtubules ($aa$), are given by
\begin{equation}
  \dot{\mathbf{x}}=\mathbf{v}^s(\mathbf x)+v_w(\mathbf{P}-\mathbf{p}),
  \textrm{  and  }
  \mathbf{\Sigma}^{aa}=\alpha_{aa}\rho(\mathcal{Q}-\mathbf{P}\mathbf{P}).
  \label{Eq:GaoFluxes}
\end{equation}
These forms for $\dot{\mathbf{x}}$ and $\mathbf{\Sigma}^{aa}$
reproduce exactly the nematic cluster results above, and show the
dependencepolarity sorting  of the microtubule velocity upon the local polarity
$\mathbf{P}(\mathbf{x})$.

This theory was further informed by detailed Brownian/Monte-Carlo simulations
of nematic assemblies of rigid filaments interacting through
multimeric kinesin-1 motors, thermal noise, and steric interactions.
From these simulations were gleaned estimates of the activity
coefficient $\alpha_{aa}$ giving that $\alpha_{aa}<0$, i.e. that
polarity sorting dipolar stresses were extensile
\cite{gao2015multiscalepolar,GBGBS2015b}. The stochastic simulations
also showed the presence of an additional subdominant, active and
extensile stress, having the form
$\alpha_{pa}\rho(\mathcal{Q}+\mathbf{P}\mathbf{P})$, arising from the
relaxation of cross-link tethers of multimeric motors connecting
polar-aligned ($pa$) microtubules \cite{blackwell2016microscopic}. With this continuum model in hand,
they analyzed and simulated the dynamics of a thin layer of active
material at the interface between two fluids. Their analysis
demonstrated the existence of a characteristic finite length-scale of
instability due to the external drag of the outer fluids, and of
orientational instabilities for aligned states. Their simulations of
the full kinetic theory showed a nonlinear dynamics of fluid and
material flows, and of nucleating/annihilating disclination defect
pairs, with a structure qualitatively similar to that observed experimentally
in \cite{sanchez2012spontaneous}.

To make comparisons to the highly cross-linked regime, the total stress
in the Gao {\it et al.} model has the form
\begin{equation}
  \mathbf{\Sigma}^{tot}=\mathbf{\Sigma}^s+\mathbf{\Sigma}^a
  =-\Pi^s\mathcal I+\mu\left(\nabla\mathbf{v}^s+\nabla\mathbf{v}^{s~T}\right)
  +(\alpha_{pa}+\alpha_{aa})\rho\mathcal{Q}
  +(\alpha_{pa}-\alpha_{aa})\rho\mathbf{P}\mathbf{P}+\tilde{\mathbf{\Sigma}}^{s}.
  \label{GaoStress}
\end{equation}
where $\tilde{\mathbf{\Sigma}}^{s}$ contains stresses arising from steric
interactions (from Maier-Saupe theory) and particle rigidity; compare to
Eq.~{(\ref{eq:stress_generic})}.
Here, the microtubule flux in Eq.~(\ref{Eq:GaoFluxes}) is of the same
form as in the highly cross-linked case but with the solvent velocity
rather than the material velocity $\mathbf{v}$
\cite{furthauer2019self}; compare Eq.~{(\ref{eq:vi_dense_limit})}. Unlike the
stress in the highly crosslinked regime, the expression in
Eq.~(\ref{GaoStress})  does not depend upon $\nabla\mathbf{P}$. It is the
relation between $\nabla\mathbf{P}$ and $\nabla\mathbf{v}$ revealed by
Eq.~(\ref{eq:self_strain_visc}) that underlays the self-straining
state of the highly cross-linked state; see Fig.~\ref{fig:fig_hydro}. 

The cluster picture in \cite{gao2015multiscalepolar} can be generalized by
allowing 'cluster-cluster' interactions which can lead to additional stresses
proportional to gradient in the local polarity. Efforts along these lines where
first made in \cite{LM2006, liverpool2008hydrodynamics}. There the key
assumption is that pairs of filaments interact via sparse coupling of constant
velocity motors. This theory
also provided an initial understanding of the origin of contractile behaviors
found in many early experiments of filament/motor mixtures.

\section{Synthesis and Open Challenges}

Cytoskeletal networks are the drivers of basic biological functions like cell
division and cell motility. The constituents of cytoskeletal networks also
provide foundational examples in the field of active matter.  In this review we
highlight the differences between the physics of highly crosslinked and sparsely
crosslinked cytoskeletal networks.
In highly crosslinked systems the long-ranged transport of momentum is mediated
through
the crosslinks themselves ($\ell^g \gg \ell^s$),
whereas in sparsely crosslinked systems it is the solvent that plays that role
 ($\ell^s \gg \ell^g$); see Fig.~\ref{fig:fig_sparse_dense}

This has important implications for the resulting material properties. 
In highly crosslinked materials, the same crosslinkers which generate the
driving forces between filaments also generate the frictional coupling that
keeps the networks coherent. Consequently, active stresses and the viscosity  of
the material are intimately linked. In particular, the polar active stress proportional to
polarity gradients is linked to the viscosity tensor by 
Eq.~{(\ref{eq:xi_related})}. 
For these materials, the viscous and polar stresses exactly balance when
all filaments in the material move at the preferred motor velocity
$\sigma_0/\gamma_0$ in the direction to which they point
\cite{furthauer2019self}. Since $\chi^v$ and
$\chi^a$ vary together the properties of this state are not sensitive to protein
concentrations. This self-straining state is  not just a theoretical curiosity;
it has been observed {\it in vitro} 
\cite{furthauer2019self} and in spindles \cite{yang08, dalton2021gelation}.

In contrast, in the sparsely crosslinked regime the viscosity tensor $\chi^v$ of the
material is largely set by the viscosity of the solvent in which filaments are
suspended and is thus independent of concentrations and properties of the
crosslinking molecules, while the values of active stresses depend on protein
concentrations. This provides a possible diagnostic for
differentiating highly crosslinked and sparsely crosslinked active materials by
modulating crosslink concentrations. Moreover, in
the self-straining state of active materials filaments can move through the
material at speeds that are independent of the local polar and nematic order
\cite{furthauer2019self}, whereas in more sparsely crosslinked systems
\cite{gao2015multiscalepolar} the motion of filaments
throughout the system depends on the local polarity.

A second striking difference between the highly crosslinked and sparsely
crosslinked materials is the form of the other active stresses $\mathcal
A^{(\mathcal Q)},\mathcal A^{(\mathbf P)}$. It has long been known that
contractile (or extensile) stresses can be generated only if motors act in a way
that breaks the symmetry between extending and contracting single filament
pairs.  This could for instance be achieved by end-clustering or end-binding
affinities of motor proteins \cite{kruse2000actively,
liverpool2008hydrodynamics}. On top of that, the mathematics of the highly
crosslinked regime requires that $\sigma_1\ne \gamma_1/\gamma_0\sigma_0$; see
Eqns.~{(\ref{eq:AQ},\ref{eq:AP})}. This means that the anisotropy of friction
($\gamma_1/\gamma_0$) along filaments needs to be different from the anisotropy
of motor drive ($\sigma_1/\sigma_0$) to generate active contractions. In
practice this can be achieved by either mixing several types of crosslinks or by
creating a crosslink with two different motor heads \cite{furthauer2021design}.
This prediction might shed insight into the well known - but so far poorly
understood - observation that in many actomyosin systems contractions only occur
if a small amount of passive crosslinker is added to the system
\cite{ennomani2016architecture}.

A third important difference between the highly and sparsely crosslinked regimes
comes from the fact that the solvent is  incompressible, while the network
itself can be compacted by active processes. As a consequence sparsely
crosslinked theories predict incompressible material flow fields, while highly
crosslinked theories can predict active bulk contraction. Thus it is tempting to
speculate that contractile systems like \cite{foster2015active,
foster2017connecting} are highly crosslinked. In contrast many classic active
nematic experiments show no clear signs of compaction
\cite{sanchez2012spontaneous, wu2017transition, chandrakar2020confinement}. 

In section \ref{sse:classify} we sought to classify 
biological and experimental active matter systems as being either highly or
sparsely crosslinked. We believe that many important
systems, such as the cell cortex and many spindles, are highly crosslinked. 
For most systems however, this
assertion remains an educated guess rather than an experimental
certainty. Given the increasing awareness that the physics of active
cytoskeletal networks can be drastically different for different numbers of
crosslinks, we hope that this review will serve to highlight the need for
experiments that will answer this question more definitively.

Finally, we want to conclude by remarking that many active systems are likely to
live in the intermediate regime between highly and sparsely crosslinked. One
example are the microtubule asters studied in extract \cite{pelletier2020co}, which
are probably highly crosslinked in their bulk, but more sparsely crosslinked
near their boundaries. These structures are biologically important and
revelatory of new physics in an intermediate crosslinking regime that is so far
barely explored. 

In summary, we firmly believe that the science of biologically active materials
will benefit greatly by simply enumerating the number of crosslinks active in
the materials of interest.  This enumeration will help identify important and
distinct physical regimes, and help make theories quantitative and living
systems engineerable.

{\bf Acknowledgements:}
We thank James Pelletier, Steffanie Redemann, and Kun-Ta Wu for providing us
with experimental images. We thank Lucy Reading-Ikkanda for
preparation of figures.  MJS acknowledges support from NSF grants DMR-2004469
and DMR-1420073 (NYU-MRSEC). We thank our many collaborators who worked with us
in this area.

\bibliographystyle{unsrt}

\end{document}